\documentclass[%
 twocolumn,
 amsmath,amssymb,
 aps, physrev,
]{revtex4-2}

\usepackage{caption}
\usepackage{appendix}
\usepackage{graphicx}
\usepackage{dcolumn}
\usepackage{tabularx}
\usepackage{bm}

\begin{document}

\preprint{APS/123-QED}

\title{\textbf{Optimising Re-entrant Cavity Designs for Low Mass Axion Haloscopes} 
}%

\author{Raj Aryan Singh$^{1}$}
\email{Contact author: rasingh@swin.edu.au}
\author{Paige Rose Taylor$^{1}$}
\author{Elrina Hartmann$^{2}$}
\author{Geoffrey Brooks$^{1}$}
\author{Ben T. McAllister$^{1,2}$}

\affiliation{%
$^{1}$Centre for Astrophysics and Supercomputing, Swinburne University of Technology, John St, Hawthorn VIC 3122, Australia
}

\affiliation{%
$^{2}$Quantum Technologies and Dark Matter Laboratory, University of Western Australia, Crawley WA 6009, Australia
}

\date{\today}

\begin{abstract}
Axion haloscopes provide a leading experimental approach to detecting QCD axion dark matter through resonant axion–photon conversion in microwave cavities. Extending haloscope sensitivity to low axion masses remains challenging due to the large resonator volumes required at sub-GHz frequencies. Re-entrant cavities offer a compact solution, but their performance depends strongly on geometric optimisation. We present a comprehensive finite-element study of re-entrant cavity haloscope designs operating in the $100$ to $500~\mathrm{MHz}$ range, comparing their performance using effective scan time as a figure of merit. Among the configurations studied, we identify a ``double attack'' geometry that achieves a roughly threefold improvement in effective scan time compared to a conventional single-rod re-entrant cavity. We further investigate practical implementation strategies, including a hybrid design employing one fixed and one tunable rod, which preserves a scan time gain while reducing mechanical complexity. These results demonstrate a pathway to enhanced low-mass axion haloscope sensitivity.


\end{abstract}

\maketitle
\section{\label{sec:level1}Introduction}
Dark matter \citep{DM} is a cosmic enigma that accounts for $85 \%$ of the total matter of the universe. It does not interact with baryonic matter, making it invisible to telescopes. However, its presence is inferred from its gravitational effects on visible matter such as deviations in galaxy rotation curves \citep{verarubin} to galaxy cluster dynamics-and gravitational lensing \citep{bullet} consistent with general relativity. In the standard cosmological model ($\Lambda$CDM) \citep{LCDM}, dark matter (DM) is crucial for the formation and evolution of large-scale structures. Theoretical candidates for DM cover a broad range of possibilities. Axions \citep{PQAXION, AXION2} are one of the most compelling dark matter candidates. They were introduced as a solution to the strong CP problem \citep{cppseudo, ansonhook}, which in simple terms is the non-observance of electric dipole moment of the neutron \citep{neutronedm}. The theoretically predicted properties of axions match the properties a desirable dark matter candidate \citep{PhysRevLett.50.925}, however, its mass cannot be predicted theoretically. Some constraints on axion mass have been placed by various astrophysical observations and direct detection experiments \citep{AxionLimits}. Despite these constraints, the axion mass search range spans several orders of magnitude from. In addition to this, the axion particles couple \citep{semertzidis2022axion} very weakly to the standard model particles, with interaction Lagrangians which can be categorized into three types : 

$$g_{a\gamma \gamma}a\textbf{E}.\textbf{B} , \ g_{aff}\nabla a.\hat{\textbf{S}}, \ g_{\text{EDM}}a.\hat{\textbf{S}}.\textbf{E}$$
    
The interactions shown above are proportional to electromagnetic fields \citep{EMF1, HAYSTAC}, fermion spins \citep{QUAX, ARIADNE} and nuclear electric dipole moment (EDM) \citep{casperspin} respectively. The first type of interaction is often probed with resonant cavities placed in a strong magnetic field, a class of experiment known as the axion haloscope \citep{sikivie}, and is what we shall focus on here.

The typical axion haloscope works on the principle that the incoming axion is expected to interact with the sea of virtual photons provided by a strong magnetic field and convert to real photons, with a frequency dependent on the axion mass and velocity:
\begin{eqnarray}
    h\nu_a = m_ac^2 \ + \ \frac{1}{2}m_av_a^2,
\end{eqnarray}
where $\nu_a$ is the frequency of the generated real photon, $m_a$ is the axion mass and $v_a$ is the axion velocity which is taken to be $\sim 10^{-3}c$ in common cold dark matter models\citep{sikivie}. The velocities of cold dark matter axions are expected to have a Maxwell-Boltzmann distribution, causing a narrow effective linewidth of the induced axion-photon signal. If the frequency $(\nu_a)$ of a generated photon matches the resonant frequency of a suitably axion sensitive resonant mode of the cavity, the photons can excite the cavity mode and can be measured as excess power above the expected background of the system.

The power deposited \citep{simanovskaiapower, MIT} due to the resonant conversion of axion to photon is given by:
\begin{equation}
\begin{split}
P_{signal}  = \left(g_{a\gamma \gamma}^2 \frac{\hbar^3 c^3 \rho_a}{m_a^2} \right)\\
\times 
\left(\frac{1}{\mu_0}B_0^2\omega_c V C_{nml}Q_L \frac{\beta}{1+\beta}\frac{1}{1+(2\Delta\nu_a / \Delta\nu_c)^2} \right)
\end{split}
\label{eq:Psig}
\end{equation}

where the first pair of parentheses contain universal constants like the speed of light $c$, reduced Planck's constant $\left (\hbar = \frac{h}{2\pi}\right )$ and the unknown parameters of axion and dark matter theory -- like the axion-photon coupling constant $g_{a\gamma \gamma}$, axion mass $m_a$, and local dark matter density $\rho_a$ \citep{LocalHaloDensity}. The second pair of parentheses contain experimentally controlled parameters like the applied magnetic field $B_0$, available cavity volume $V$, cavity resonant frequency $\omega_c = 2\pi\nu_c$, loaded quality factor of cavity $Q_L$, a mode-dependent form factor $C_{nml}$ \citep{mcallister2017higher}, cavity-antenna coupling $\beta$. $\Delta \nu_a$ is the axion linewidth, $\Delta\nu_c$ is the cavity linewidth and $\mu_0$ is the permeability of free space. The form factor $C_{nml}$ quantifies the overlap of the electric field of the resonant mode and the external magnetic field $B_0$. It is defined as:
\begin{eqnarray}
    C_{nml} = \frac{\left( \int \bm{E}.\bm{B_0}dV \right)^2}{B_0^2V\int |\bm{E}|^2dV}
\end{eqnarray}
In a typical haloscope, the power in the cavity is measured by some form of linear amplification (although next generation experiments seek to move towards single photon counting). In the linear amplification case, the signal-to-noise ratio in an axion haloscope is given by:
\begin{eqnarray}
    SNR = \frac{P_{signal}}{k_BT_{sys}} \sqrt{\frac{\tau}{\Delta\nu_a}}
    \label{eq:snr}
\end{eqnarray}
where $\tau$ is the integration time, $k_B$ is the Boltzmann constant, $T_{sys}$ is the system noise temperature -- not necessarily the same as physical temperature of the system, but including contributions from the amplification chain.

\section{Low Mass Haloscopes and Re-entrant Cavities}

A major challenge in the search for axions is the non-predictability of the axion mass, which necessitates scanning a vast parameter space. Although experiments like ADMX \cite{ADMX,PhysRevLett.134.111002}, ORGAN \citep{ORGAN, Quiskamp_2022, quiskamp2025near,ORGAN1B}, and many others have successfully designed tunable resonant cavities to operate in GHz range, it is challenging to cover the lower frequency range due to the inverse relationship between cavity frequency and cavity size, meaning that at lower frequencies, cavities tend to become very large, and this presents a challenge for insertion into high field magnet bores. Re-entrant cavities \citep{BEN_LC} provide a practical way to scan this lower end of frequency range without requiring impractically large resonators. A re-entrant cavity can be conceptualised as a lumped LC circuit where the electric and magnetic fields are spatially separated and highly localised. A standard re-entrant design has a main cylindrical cavity and a tuning rod that can be moved in and out of the cavity along the central axis of the cavity. This movement of the tuning rod inside the cavity pushes the resonant frequency down which can be easily explained using the equivalent lumped LC circuit for the haloscope. Re-entrant cavities have been discussed as a possible method for accessing the low mass axion range with haloscopes, and some work has been done to investigate various designs \citep{LOWFREQ}. However, much room remains for designs to be optimised further, which is the goal of this study.

\section{Haloscope Figure of Merit and Comparison Methodology}

Given that a haloscope must be tunable over a range of frequencies, the overall figure of merit accepted by the community is the allowable rate of frequency scanning at a given level of sensitivity. The scan rate $S$ can be derived by substituting Eq.~\ref{eq:Psig} in Eq.~\ref{eq:snr}
\begin{equation}
\begin{split}
S \equiv \frac{d\nu}{dt} \approx \frac{4}{5} \frac{Q_L Q_a}{(SNR)^2} 
\left( g_{a\gamma \gamma}^2 \frac{\hbar^3 c^3 \rho_a}{m_a^2} \right)^2 \\
\times 
\left( \frac{1}{\hbar\mu_{0}} \frac{\beta}{1+\beta} B_0^2 V C_{nml} \frac{1}{N_{sys}} \right)^2
\end{split}
\label{eq:scanrate}
\end{equation}

where $N_{sys}$ is the effective system noise photon number \citep{simanovskaia2019design}, $Q_a$ is the axion quality factor $(\sim 10^6)$ \citep{axionQ}, and other variables have the same meaning as defined above. For a given set of fixed experimental parameters like $SNR$, $B_0$, $\beta$, $N_{sys}$ and other theoretical parameters, the scan rate is dependent on the cavity design, and is proportional to $C^2V^2Q$. 
\begin{eqnarray}
    \frac{d \nu}{dt} \propto C^2V^2Q
    \label{eq:cvq}
\end{eqnarray}
From here on, we will use $Q$ \& $Q_L$ interchangeably unless otherwise stated. 

\begin{figure*}[t]
    \centering
    \includegraphics[width=0.75\linewidth]{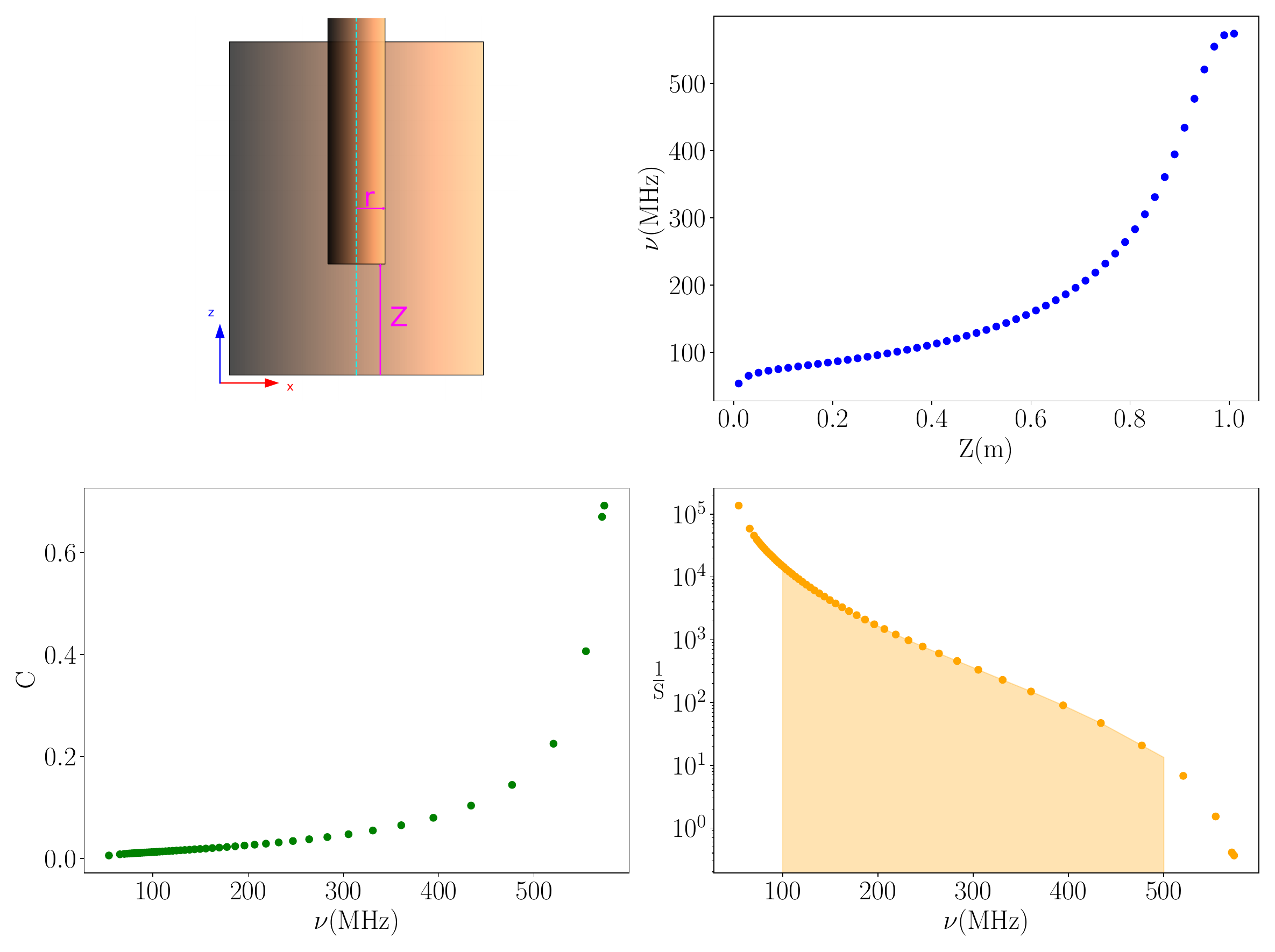}
    \caption{(Top left) Geometry of the single re-entrant tuning rod, showing the rod radius $r$ and the tuning parameter $\mathrm{Z}$ defined as the axial displacement of the rod. The dashed line in cyan is the central axis of symmetry for the setup. (Top right) Resonant frequency $\nu$ as a function of the tuning parameter $\mathrm{Z}$ tuned at a step size of $2 \ \mathrm{cm}$. (Bottom left) Cavity form factor $\mathrm{C}$ as a function of resonant frequency. (Bottom right) Inverse scan rate $\mathrm{1/S}$ as a function of frequency, with the shaded region indicating the contribution to the total scan time relation.}
    \label{fig:SINGLE ROD REENTRANT}
\end{figure*}

\begin{eqnarray}
    T = \int_0^{T}dt \propto \int_{\nu_1}^{\nu_2}\frac{1}{C^2V^2G}d\nu
\end{eqnarray}

For ease of computation in comparing different cavity designs, it is advantageous to replace $Q$ with $G$ in Eq~\ref{eq:cvq}, which is a purely geometric factor defined as:
\begin{eqnarray}
     G = \frac{2\pi \nu\mu_{0} \int |\textbf{H}|^2dV }{\int|\textbf{H}|^2dS}
     \label{eq:G}
\end{eqnarray}
and can be related to the quality factor of the cavity by \citep{QRG}: 
\begin{eqnarray}
    Q = \frac{G}{R_s}
    \label{eq:QG}
\end{eqnarray}
where $R_s$ is the surface resistance of the material of resonant cavity. Thus Eq.~\ref{eq:cvq} can now be written as:
\begin{eqnarray}
    \frac{d\nu}{dt} \propto C^2V^2G
\end{eqnarray}
where we have dissolved $R_s$ in the proportionality constant as it's sole role is in the material choice which can be fulfilled by choosing a material that is a good conductor of electricity. 

For the purpose of comparing different cavity designs, we define an effective `scan-time' which is proportional (up to the fixed parameters above) to the time it would take to scan a range of axion parameter space at a fixed sensitivity level. The scan time can be defined as the integral of the scan rate over the given frequency range. 

\section{Comparing Designs}
Having defined the effective scan-time relation as a quantitative tool for comparing the performance of different designs, we turn our attention to the re-entrant cavity designs that we will compare. 
For the purpose of comparison to previous work \citep{BEN_LC, LOWFREQ}, and to have a standard size for comparison across designs, a cylindrical cavity with radius $R = 0.2$ m and height $H= 1$ m is chosen as the outer cavity.
We consider a variety of different re-entrant rod configurations inserted into this outer cavity, simulate them using finite element analysis in COMSOL Multiphysics, extract the relevant cavity parameters, and integrate the inverse of the scan rate of the frequency range of interest (chose to be $100$ to $500 \ \mathrm{MHz}$ in this study) to find the effective scan time. 
First, both as a baseline and a validation of the analysis method, we consider the simple single cylindrical rod re-entrant cavity, and the telescope rod cavity discussed in previous work \citep{BEN_LC, LOWFREQ}. These designs form the baseline for comparison with the new designs studied. Recalling basic electrostatics, sharp corners imply high charge concentration and eventually strong electric fields at these corners - this intuition forms the motivation for several of the designs to be rigorously studied.

\begin{figure*}[t]
    \centering
    \includegraphics[width=0.75\linewidth]{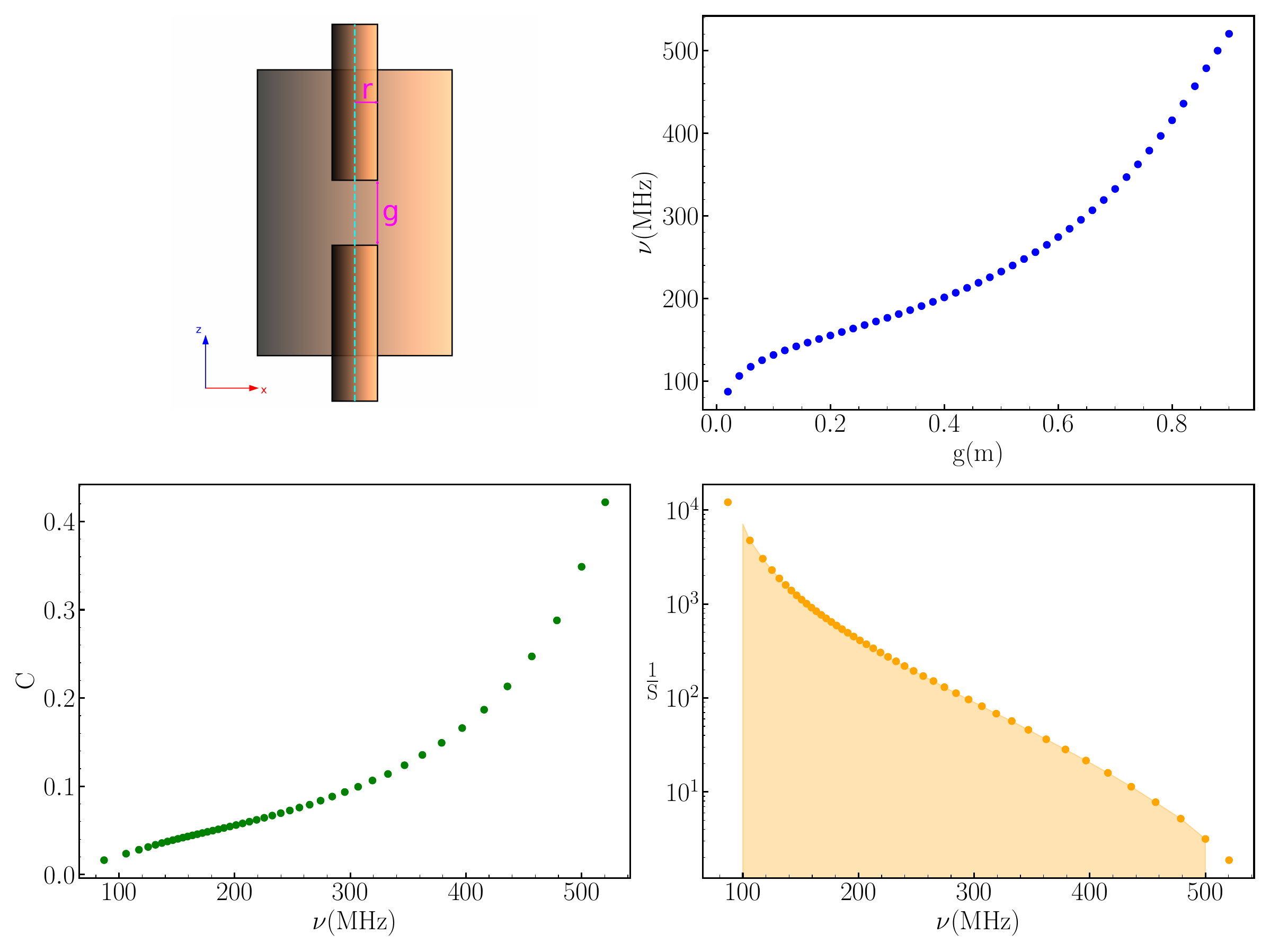}
    \caption{(Top left) Schematic of the double attack cavity design, showing two tunable rods inserted from the opposite ends. The parameters $r$ and $\mathrm{g}$ denote the rod radius and inter-rod gap, respectively. The remaining panels follow the same layout and definitions as in Fig.~\ref{fig:SINGLE ROD REENTRANT}}
    \label{fig:DA_C}
\end{figure*}

\subsection{Benchmark: Single Rod}
This is the first benchmark design for our comparison. As shown in Fig.~\ref{fig:SINGLE ROD REENTRANT}, the design features a tuning rod that can be moved along the $z$ axis to tune the cavity and the $Z$ parameter defines the position of the tuning rod with respect to bottom of the cavity cylinder. In our study, we swept over the radius $r$ of the tuning rod to find the optimum radius, i.e. the tuning rod radius for which the design has the minimum scan time relation, which was found to be $r = 0.09 \ m$ with $T = 6.49 \times 10^5$. Note that this value of $r$ is $45 \%$ of $R$ where $R$ is the radius of the cavity \citep{LOWFREQ}. This is in agreement with previous work, and forms the scan time which future designs must beat.

\subsection{Second Benchmark: Telescopic Rod}

The telescopic post design was originally considered in \citep{LOWFREQ}, drawing on the intuition that more sharp corners will concentrate field, and thus boost form factor. The design does offer enhanced scan time compared to the simple post, but comes with the cost of increased mechanical complexity. Regardless, we compare it to our new designs as another benchmark to beat.

The design we studied consisted of four segments with radii $r_1 , r_2, r_3, r_4$  as shown in Fig.~\ref{fig:telescopic_C} (see Appendix~\ref{app:plots}). We swept over the radii of all the segments, and found the optimum scan time when $r_1 = 0.09 \ m$ and $r_i = r_{i-1} + 0.02 \ m$ where $i$ can take values $2 , 3, 4$. The scan time relation for this design was $5.47\times 10^5$ which is $\sim15\%$ better than the standard re-entrant design.

However, as noted, this design comes with a high possibility of mechanical complications -- the segments can exhibit tilting and sticking during deployment, complicating precise control over the tuning process. The practical difficulties of this design, despite its improved sensitivity compared to the simple design, motivated the further designs studied here. We search for a mechanically simpler design with a sensitivity beating the simple post design, and ideally beating the telescopic design as well.

\subsection{Dual Rod}
The first new design idea was motivated by the desire to have more sharp corners, whilst simplifying the mechanical design compared to the telescope rod. We tested the dual rod design, shown in Fig.~\ref{fig:DUALROD} (see Appendix~\ref{app:plots}). The designs consists of two parallel cylindrical rods, each with radius $r$ at a separation $d$ between them, that can move along the $z$ axis and $Z$ is the position of the tuning rod with respect to the bottom of the cavity. We swept over a range of combinations of $r$ and $d$ to find the optimum values, which we found to be $r = 0.05 \ m$ , $d = 0.07 \ m$, resulting in an effective scan time over the frequency range of interest of $T = 8.66 \times 10^5$. This is a higher scan time than both of our benchmarks, making it an unattractive design in our desired frequency range.

\begin{figure*}[t]
    \centering
    \includegraphics[width=0.95\linewidth]{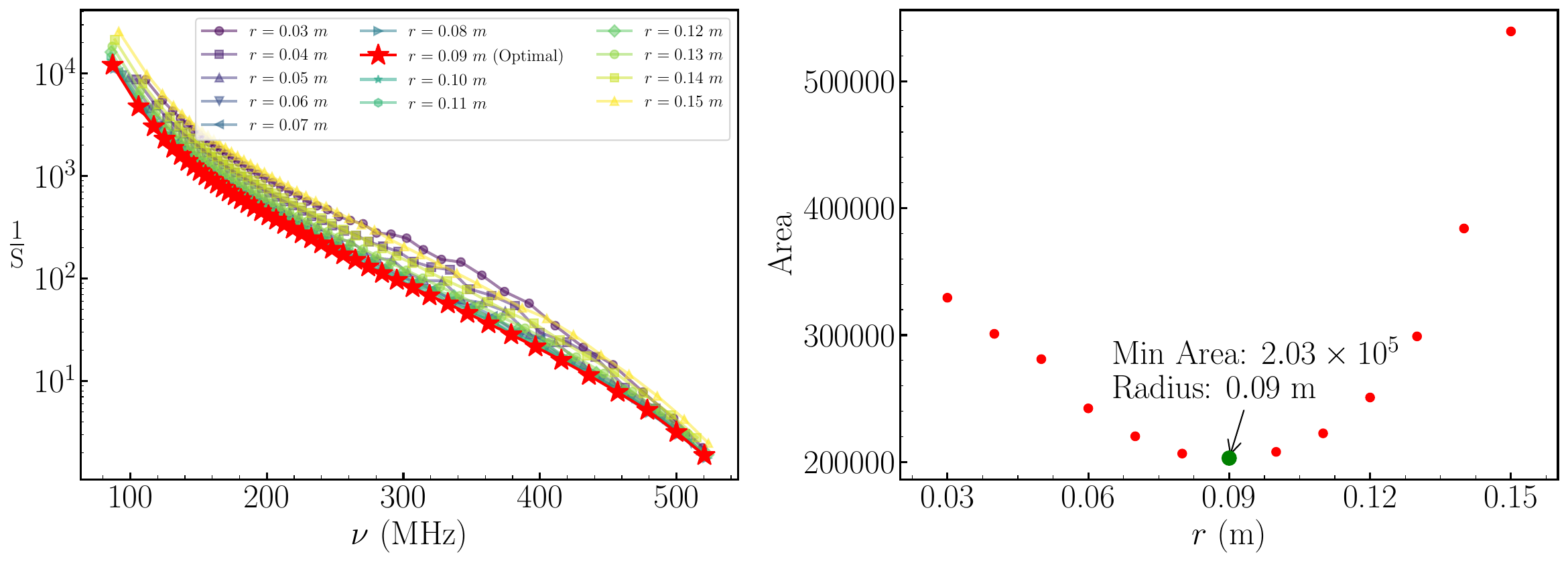}
    \caption{(Left) Inverse scan rate ($\mathrm{1/S}$) as a function of frequency $\nu$ for a range of tuning rod radii ($r = 0.03$ to $0.15$ m) for the double attack design. The configuration with $r = 0.09$ m, highlighted by the red stars, consistently yields the minimum $\mathrm{1/S}$ values, identifying it as the optimal geometry for minimising scan time across the $100–500 \ \mathrm{MHz}$ frequency range. (Right) Area under the curve (scan-time relation) of $\mathrm{1/S}$ vs $\nu$ plotted against radii $r$ of tuning rods, confirming the optimum performance at $r = 0.09 \ \mathrm{m}$}
    \label{fig:DA_ARCS}
\end{figure*}

\subsection{`Spiky' Rod}
In this study, we introduce a series of spikes on a single cylindrical tuning rod to boost the form factor due to enhanced electric field concentration at the spikes. We tested a `spiky' rod design consisting of $n$ spikes of equal size situated atop the tuning rod (Fig.~\ref{fig:SPIKEYBEST} in Appendix~\ref{app:plots}). We fixed the radius of the tuning rod to be $r_t = 0.09 \ m$ as it was found to be the best radius for the single rod design. We then swept over the number $n$ (and radius $r_s = r_t/n$) of the spikes, and the height of each spike $h_s$. Again, the rod can be tuned by varying the parameter $Z$. The minimum scan time obtained was $6.17 \times 10^5$ for $n = 2$ and $h_s = 0.08 \ m$. This performance is inferior to that of the telescopic rod design, while offering only a marginal improvement of about $5 \%$ over the single rod (benchmark) design.

\subsection{Polygonal Rod}
Instead of introducing spikes on the tuning rod, we investigate an alternative approach in which the cross-sectional geometry of the tuning rod is modified. Specifically, the tuning rod was modelled with a uniform $n$-sided regular polygonal cross section, whose vertices lie on a circle of radius $R'$. The edge length $a_n$ for such $n$-sided regular polygon is given by $2R'\sin(\frac{\pi}{n})$. For polygons with $n = 3,4,5,...,9$, we performed a parameter sweep over $R'$ to determine the optimum value of $R'$ for each geometry and to identify the best polygonal design (Figs.~\ref{fig:POLYGONS},~\ref{fig:HEXROD} in Appendix~\ref{app:plots}). None of the polyongal designs were found to outperform the benchmark design. This lack of improvement is attributed to the fact that as $n$ increases, the polygonal cross section asymptotically approaches a cylinder of radius $R'$. This convergence was verified by a simulation with $n = 40$ which yielded a scan time of $6.51 \times 10^5$ (only about $0.25 \%$ off from single rod scan time). Furthermore, for $n \geq 8$, the optimal value of $R'$ converges towards $r = 0.09 \ m$, which corresponds to the optimum radius of tuning rod for the single rod design.

\begin{figure*}[t]
    \centering
    \includegraphics[width=0.75\linewidth]{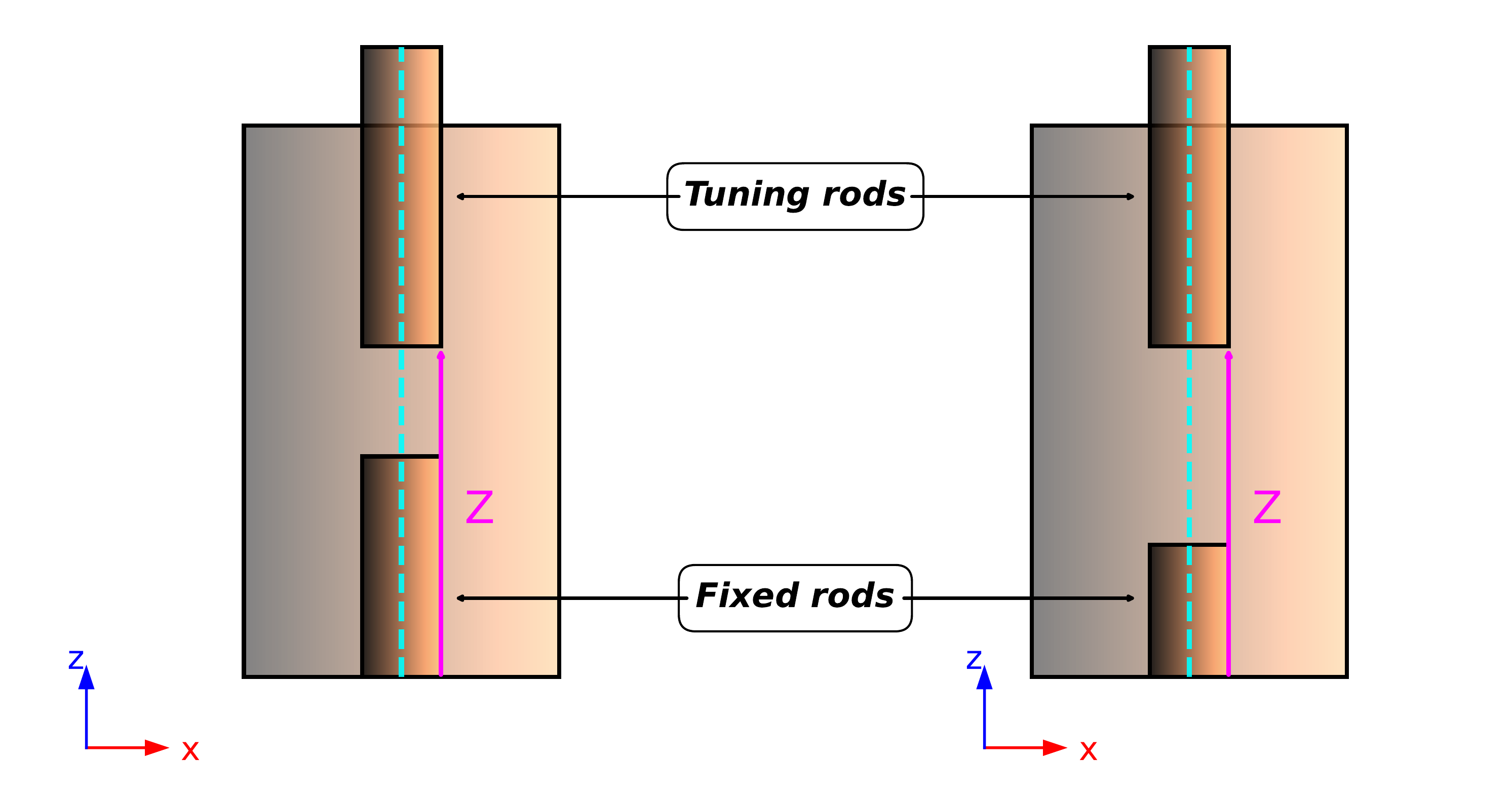}
    \caption{Schematic of hybrid double attack design showing the primary fixed rod (left) and the secondary fixed rod (right), along with the tuning rod and the tuning parameter $Z$.}
    \label{fig:onefix}
\end{figure*}

\subsection{Double Attack}

Finally, we tested the `double attack' design (Fig.~\ref{fig:DA_C}). This design extends the standard single rod re-entrant design by incorporating two tuning rods of radius $r$, positioned at opposite ends of the cavity and tuned simultaneously to maintain the system's symmetry relative to the central plane passing perpendicular to the longitudinal axis. The possibility of improving the conventional single-rod re-entrant cavity by introducing a second rod from the opposite end was briefly noted in previous work. Motivated by this observation, we independently perform a detailed and systematic study of this configuration, which we refer to as the “double attack” design, and demonstrate its substantial advantages for low-mass axion haloscopes. We define the gap between two rods as $g$, which is functionally equivalent to the parameter $Z$ used in our other designs. A parametric sweep over values of $r$, identified the optimal radius to be $r = 0.09 \ m$ (Fig.~\ref{fig:DA_ARCS}). The scan time-relation for this design was calculated to be $2.03\times 10^5$, representing a more than threefold improvement over the standard re-entrant design.

\begin{figure*}[t]
    \centering
    \includegraphics[width=0.75\linewidth]{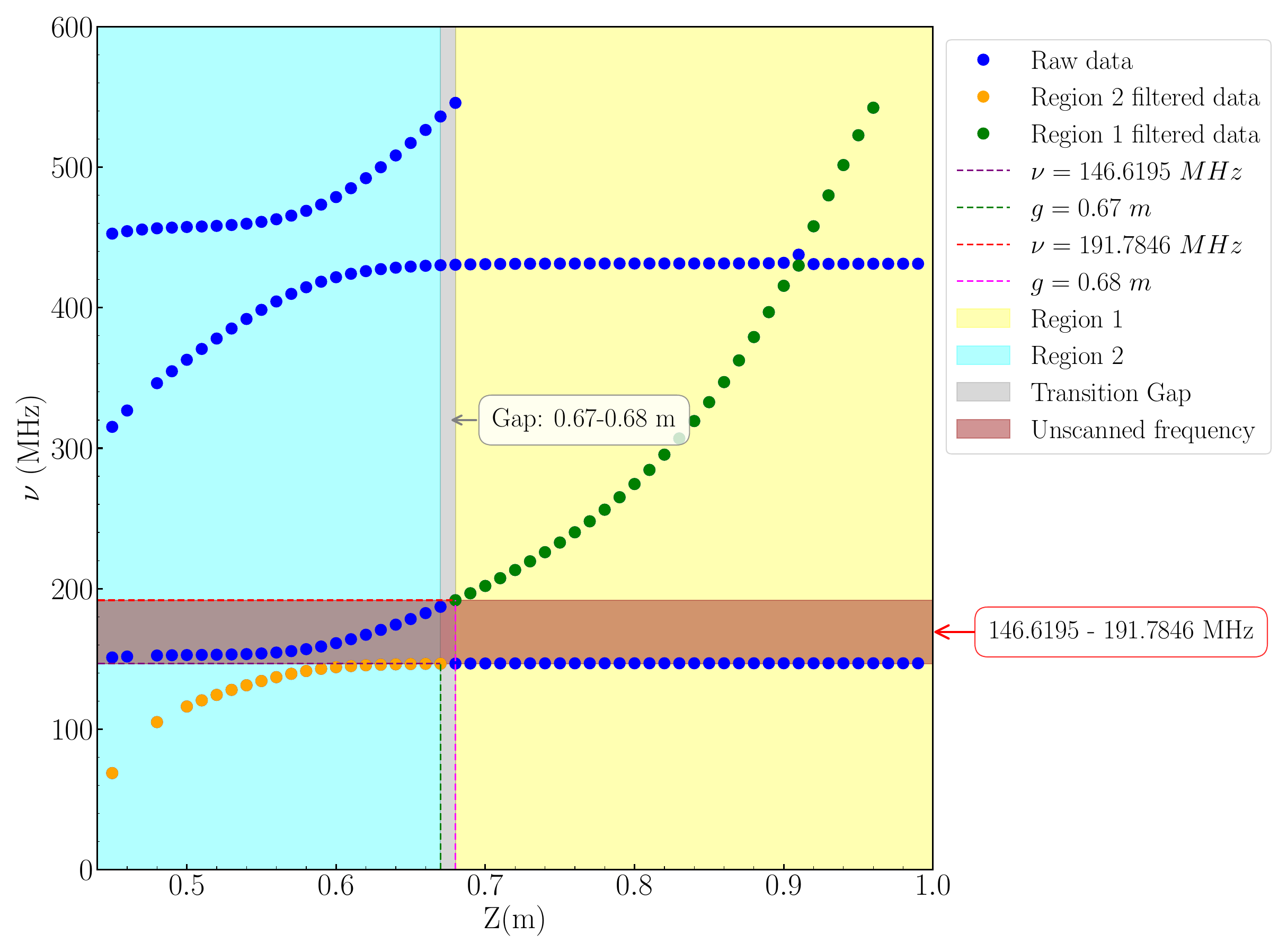}
    \caption{Resonant frequency $\nu$ as a function of the parameter $\mathrm{Z}$ when using a fixed rod of height $h = 44 \ \mathrm{cm}$. A clear interaction between the axion-sensitive tuning mode and non-sensitive intruder mode is observed, resulting in an avoided level crossing. This interaction creates an un-scannable frequency band of approximately 46 MHz (shaded brown region), spanning from 146.6 MHz to 191.8 MHz. The filtered data in Regions 1 (yellow) and 2 (cyan) highlight the operational tuning range, while the transition gap ($\mathrm{Z} \approx 0.67$–$0.68$ m) marks the beginning of mode mixing zone where sensitivity is significantly degraded.}
    \label{fig:44plot}
\end{figure*} 

\begin{table}[h]
\caption{Table 1 : Scan Time relations across different designs \label{tab:table1}}
\begin{ruledtabular}
\begin{tabular}{cc}
 {Design Name} & {Time Relation}\\
\hline
Single Rod & $6.49 \times 10^5$ \\
Telescopic Rod & $5.50 \times 10^5$ \\
Dual Rod & $8.66  \times 10^5$ \\
`Spiky' Rod & $6.17 \times 10^5$ \\
Polygonal Rod & $6.51 \times 10^5$ \\
Double Attack & $2.03 \times 10^5$ \\
\end{tabular}
\end{ruledtabular}
\end{table}

\section{Practical Implementation of Double Attack}
The double attack design is a highly promising design, offering the best scan time of all designs studied, while also substantially reducing the mechanical complexity compared to the telescopic rod design. However, the requirement of two independent tuning rods necessitates synchronised tuning mechanisms, and the need to accommodate the two tuning rods outside the cavity, under the tight spatial constraints imposed by the magnet bore and cryogenic systems. 

To address these limitations, we propose a hybrid cavity configuration consisting of a fixed rod of a specific height $h$ mounted at the base of the cavity and a single tuning rod inserted from the opposite end, as shown in Fig.~\ref{fig:onefix}. In order to determine the impact of this modification on the performance of the double attack design, the hybrid cavity configuration was simulated for a range of fixed rod heights. The optimal fixed rod height was found to be $h=49$ cm, for which the scan time relation is $2.78 \times 10^5$, which is only a $\sim37\%$ increase, and still nearly two times faster than the single rod design. 

However, the presence of the fixed rod introduces some complications. The fixed rod supports the presence of some non-axion-sensitive `intruder' modes at fixed frequencies. When the resonant frequency of the tuning mode (axion sensitive mode) is close to the resonant frequency of these non-sensitive mode, avoided level crossings can occur, degrading the sensitivty of the tuning mode to axions, and inhibiting the ability of the detector to scan a small frequency range. Fig~\ref{fig:44plot} demonstrates the mode interaction in the lower end of the frequency range. Different fixed rod heights lead to different un-scannable bands, as summarised in Table 2.   

\begin{table}[t]
\begin{ruledtabular}
\caption{Table 2 : Performance across different fixed rod heights\label{tab:table2}}
\begin{tabular}{ccc}
 {Rod height $\mathrm{(cm)}$} & {Time relation} & {Un-scanned band $\mathrm{(MHz)}$}\\
\hline
49 & $2.78 \times 10^5$ & $\sim 133$ - $187$ \\
44 & $2.88 \times 10^5$ & $\sim 146$ – $192$ \\
39 & $3.33 \times 10^5$ & $\sim 162$ – $197$ \\
34 & $4.38 \times 10^5$ & $\sim 182$ – $208$ \\
29 & $5.35 \times 10^5$ & $\sim 207$ – $226$ \\
\end{tabular}
\end{ruledtabular}
\caption{For each fixed rod height, the table lists the range of the unscannable frequency band and the corresponding scan time relation for the accessible frequency range.\label{tab:table1}}
\end{table}

If such un-scanned bands are unacceptable to a given experiment, they can be mitigated by temporarily replacing the primary fixed rod with a secondary fixed rod of a different height (Fig.~\ref{fig:onefix}). By selecting a secondary rod height that shifts the mode interaction to a different frequency range, the previously in-accessible band can be scanned.

To demonstrate this strategy, scan times obtained for several primary fixed rod heights (Table 2) were combined with the additional scan time required to cover the corresponding un-scannable frequency bands using a secondary fixed rod. The total scan time relation for the combined rod configurations are presented in Table 3 (see Appendix~\ref{app:plots}).

From Table 3 it can be concluded that the best configuration to scan the entire $100$ to $500 \ \mathrm{MHz}$ range is to use a 44 cm rod as the primary rod followed by a 29 cm rod as the secondary rod, yielding a total scan time relation of $4.01 \times 10^5$, which is a significant improvement over benchmark designs.

\section{Conclusion}
We have performed a systematic COMSOL-based optimisation study of re-entrant cavity haloscopes targeting the low-mass axion band corresponding to $100$ to -$500~\mathrm{MHz}$. Using the standard haloscope scan-rate figure of merit ($C^2V^2G$), we define an effective scan time relation for these designs, and use it to compare a range of rod geometries. We found the `double attack' configuration to deliver the strongest performance, achieving an effective scan-time relation of $2.03\times10^5$ - more than a factor of three faster than a conventional single-rod re-entrant cavity over the same frequency range. To address implementation constraints, we introduced a hybrid `one fixed + one tunable' variant that retains most of the sensitivity gain while reducing mechanical complexity, and we showed how unavoidable mode crossings can be mitigated by swapping fixed-rod heights to recover full frequency coverage. These results identify a clear, experimentally realistic path to substantially accelerating low-frequency haloscope searches and motivate near-term prototype development and validation in a magnet and cryogenic environment.

\section{Acknowledgements}
This work is supported by Australian Research Council Grants CE200100008 and DE250100933. We thankfully acknowledge the contributions of Aidan O'Keeffe, Zoe Ballard, and Ryan Lewandowski.

\bibliography{apssamp}

\clearpage
\onecolumngrid
\appendix
\section{}
\label{app:plots}

\begin{figure*}[h!]
    \centering
    \includegraphics[width=0.85\linewidth]{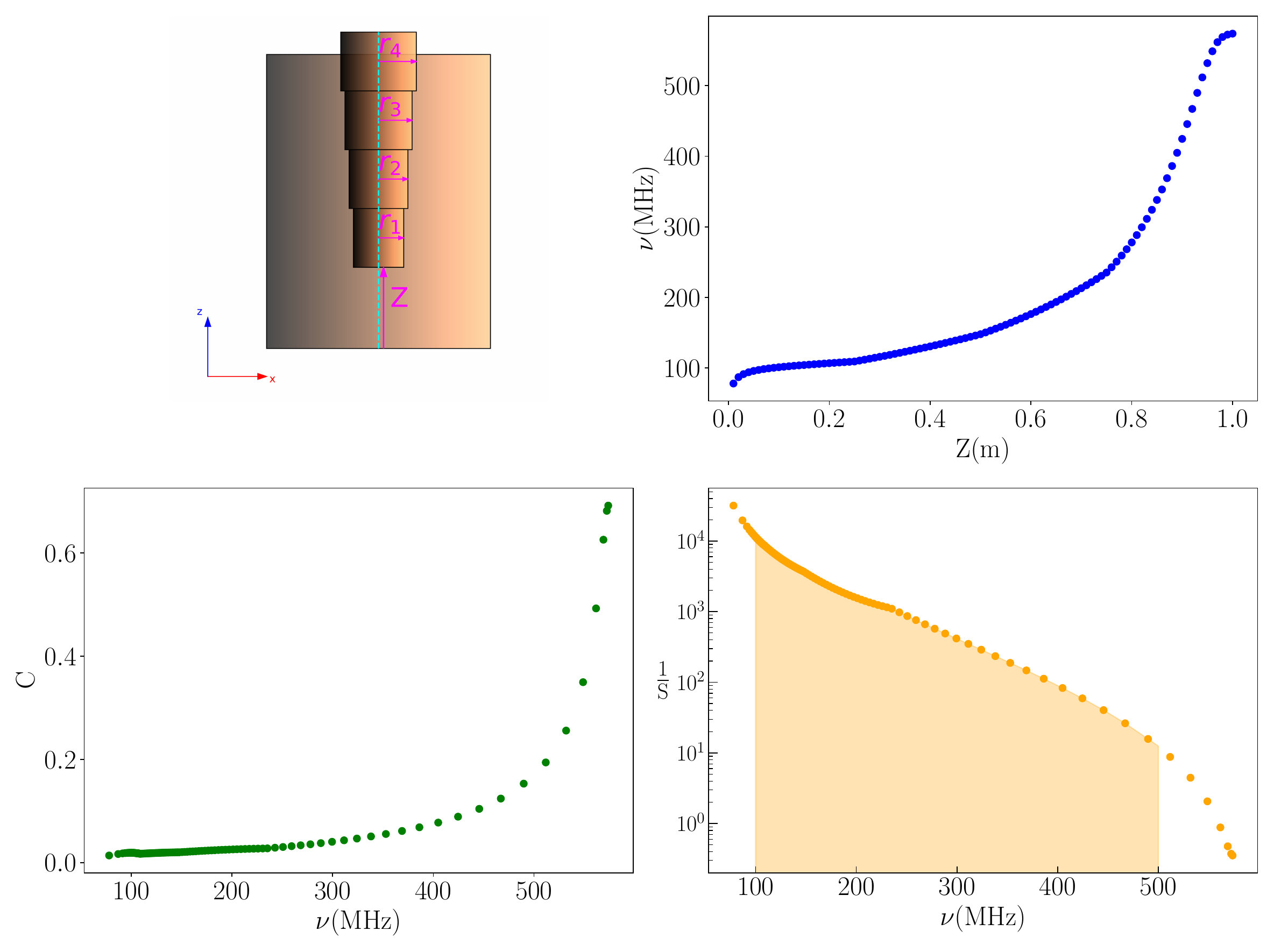}
    \caption{(Top-left): Schematic of the four-segmented telescopic rod featuring radii $r_1$ through $r_4$ and the tunable parameter $Z$. (Top-right): Tuning curve showing the resonant frequency $\nu$ of the axion sensitive mode as a function of the gap $Z$ at a step size of $1 \ \mathrm{cm}$. The remaining panels follow the same layout and definitions as in Fig~\ref{fig:SINGLE ROD REENTRANT}}
    \label{fig:telescopic_C}
\end{figure*}

\begin{figure*}[t]
    \centering
    \includegraphics[width=0.85\linewidth]{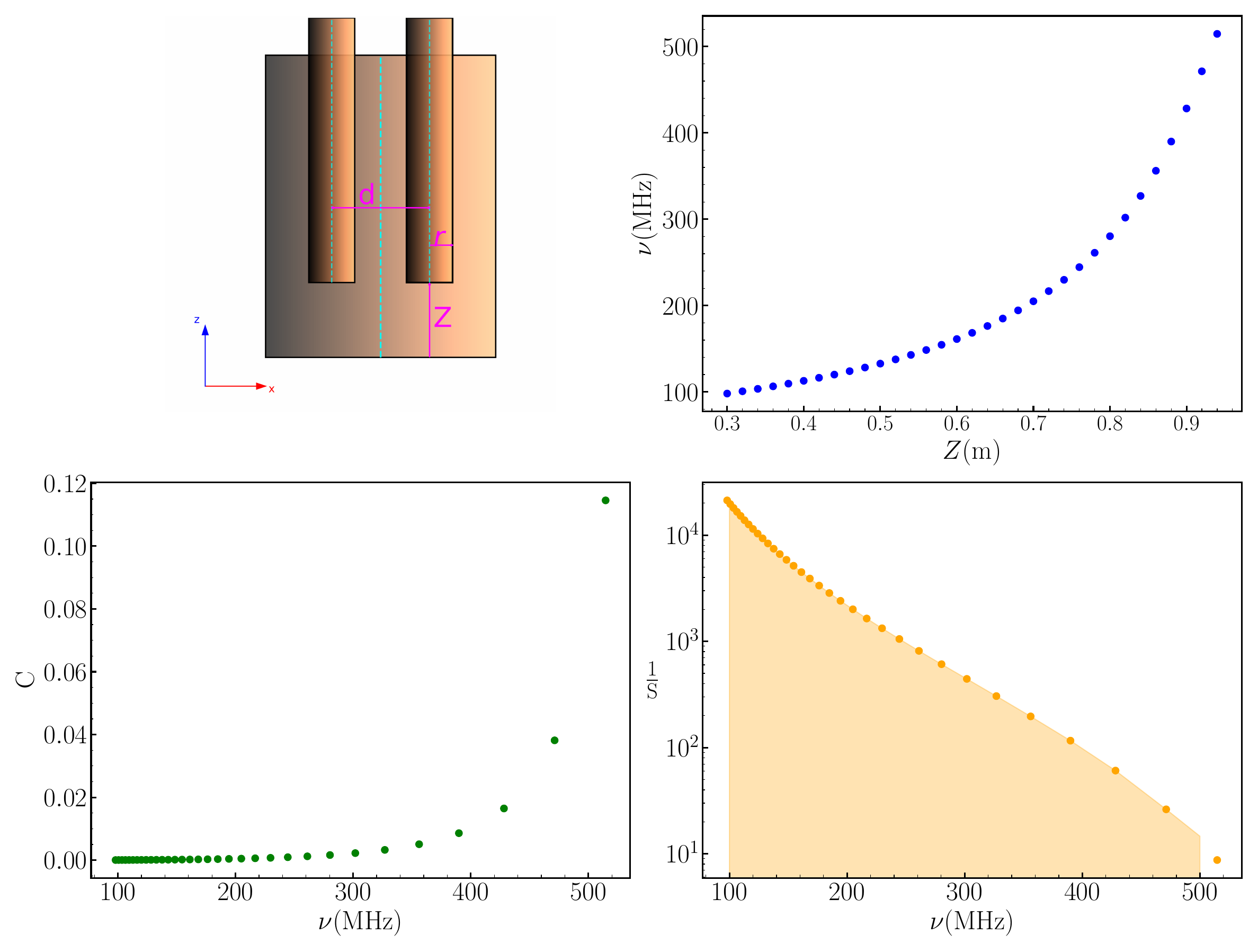}
    \caption{(Top-left): Schematic of the dual rod design showing the rod radius $r$ and the tunable parameter $Z$, along with the separation $d$ between the center of the tuning rods. The remaining panels follow the same layout and definitions as in Fig~\ref{fig:SINGLE ROD REENTRANT}}
    \label{fig:DUALROD}
\end{figure*}

\begin{figure*}[t]
    \centering
    \includegraphics[width=0.85\linewidth]{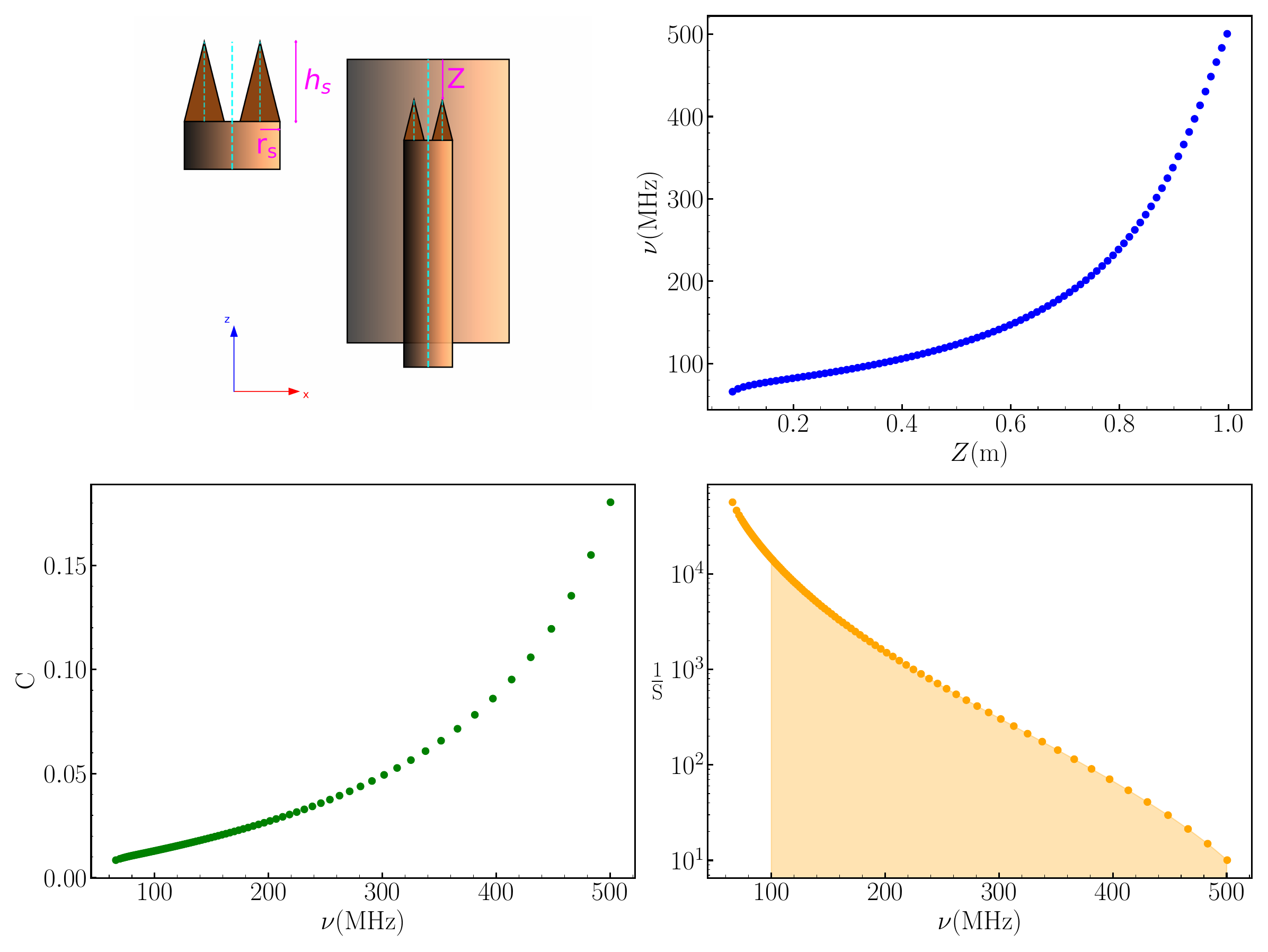}
    \caption{(Top-left): Schematic of the `spiky' rod design with the tuning parameter $Z$ and the inset shows a magnified view of the spike defining the spike radius $r_s$ and the spike height $h_s$. Note that this is the cross sectional view of the design and one can get the 3D view by revolving the setup about the central axis of symmetry (cyan dashed line running from top of the cavity to the bottom of the tuning rod). The remaining panels follow the same layout and definitions as in Fig~\ref{fig:SINGLE ROD REENTRANT}}
    \label{fig:SPIKEYBEST}
\end{figure*}

\begin{figure*}[t]
    \centering
    \includegraphics[width=0.85\linewidth]{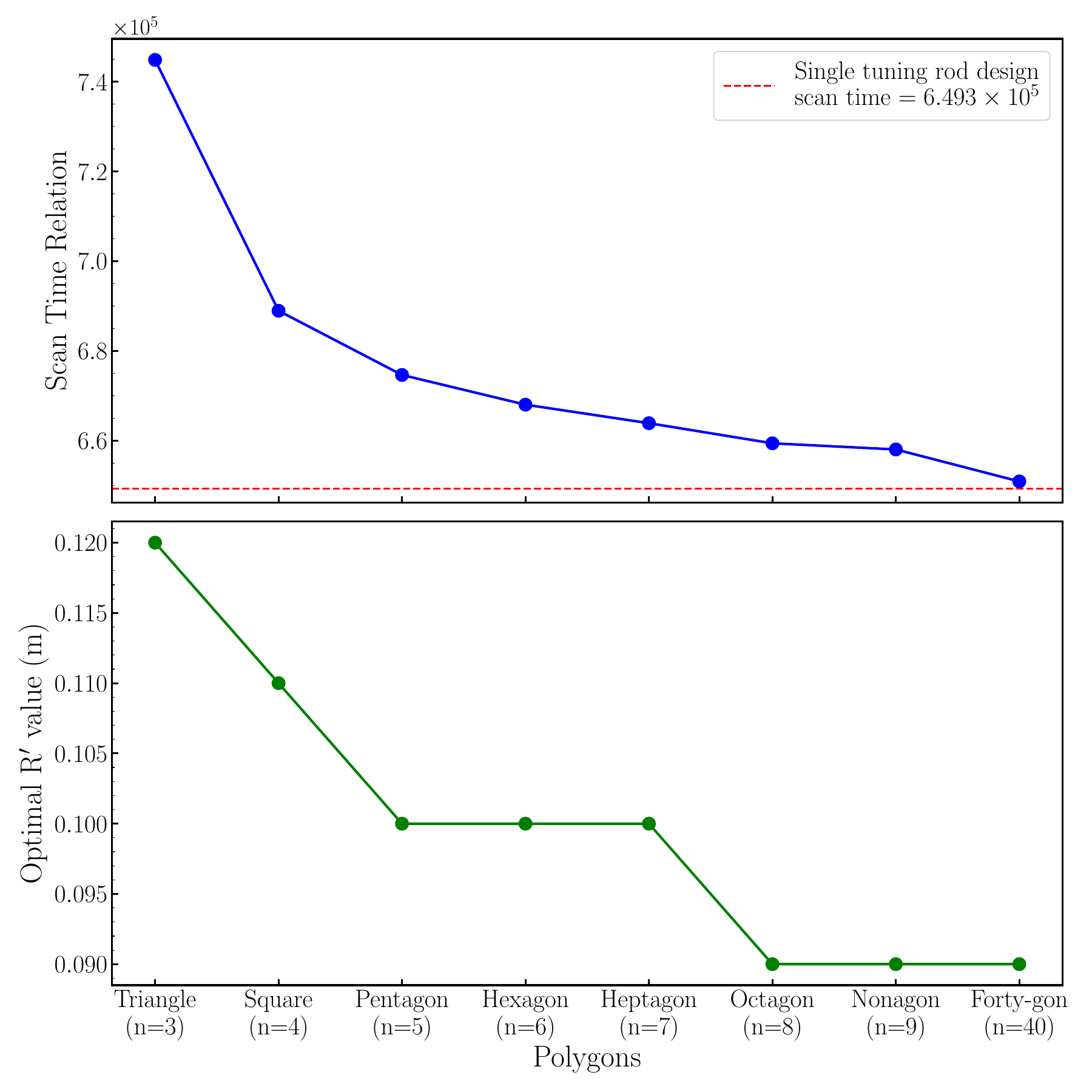}
    \caption{(Top) Scan time relation as a function of the number of polygon sides $n$. The results show that as $n$ increases, the performance asymptotically approaches the single cylindrical rod benchmark (dashed red line). (Bottom) The optimal circumradius $\mathrm{R'}$ for each geometry, showing a clear convergence toward $r = 0.09$ m for $n \geq 8$. The $n=40$ case (Forty-gon) serves as a numerical verification, yielding a scan time within $0.25 \%$ of the single rod re-entrant cavity design, confirming that polygonal modifications do not offer a performance advantage over the benchmark.}
    \label{fig:POLYGONS}
\end{figure*}

\begin{figure*}[t]
    \centering
    \includegraphics[width=0.85\linewidth]{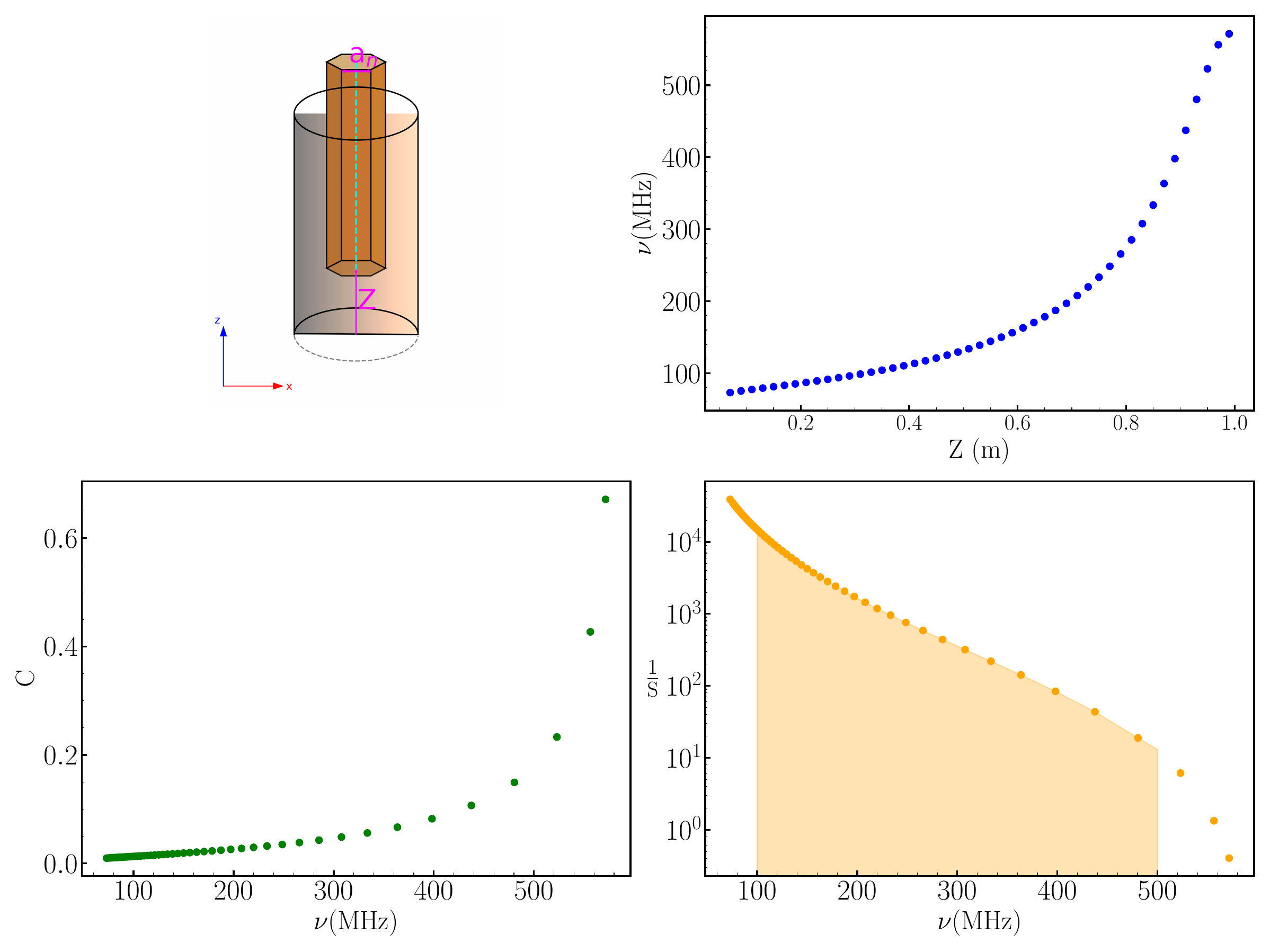}
    \caption{(Top-left): 3D schematic of the tuning rod with a regular hexagonal cross-section, defined by edge length $a_n$ and tunable parameter $\mathrm{Z}$. The remaining panels follow the same layout and definitions as in Fig~\ref{fig:SINGLE ROD REENTRANT}}
    \label{fig:HEXROD}
\end{figure*}

\clearpage

\begin{table*}[h!]
\caption{Table 3 : Hybrid Double Attack Cavity: Scan Time Contributions from Primary and Secondary Rods\label{tab:table3}}
\begin{ruledtabular}
\begin{tabular}{ccccccc}
    {$R_P \ \mathrm{(cm)}$} & {$T_{P_1}$\footnotemark[1]} & {$T_{P_2}$\footnotemark[1]} & {$R_S$ (cm)} & {Scanned range $\mathrm{(MHz)}$} & {$T_S$\footnotemark[2]} & {Total Time} \\
\hline
49 & $1.53 \times 10^5$ & $1.25 \times 10^5$ & 29 & 133 – 187 & $1.75 \times 10^5$ & $4.57 \times 10^5$ \\
44 & $1.35 \times 10^5$ & $1.53 \times 10^5$ & 29 & 146 – 192 & $1.11 \times 10^5$ & $4.01 \times 10^5$ \\
39 & $1.21 \times 10^5$ & $2.11 \times 10^5$ & 29 & 162 – 197 & $0.73 \times 10^5$ & $4.06 \times 10^5$ \\
34 & $1.00 \times 10^5$ & $3.38 \times 10^5$ & 24 & 182 – 208 & $0.44 \times 10^5$ & $4.82 \times 10^5$ \\
\end{tabular}
\end{ruledtabular}
\caption{Total scan time relations over the $100$ to $500 \ \mathrm{MHz}$ range using combinations of fixed rod heights in the hybrid double attack cavity. Each entry combines a primary fixed rod configuration with a secondary fixed rod to cover un-scannable frequency bands.}
\footnotetext[1]{$T_{P_i}$ is the scan time for $i^{th}$ region as defined in Fig~\ref{fig:44plot}}
\footnotetext[2]{$T_S$ is the scan time for S rod}
\end{table*}

\end{document}